\documentstyle[preprint,pra,aps]{revtex}
\begin{document}
\tighten
\def\question#1{{{\marginpar{\tiny \sc #1}}}}
\draft
\title{Theory of Neutrino Flavor Mixing}
\author{Christian Y. Cardall}
\address{Physics Division, Oak Ridge National Laboratory, Oak Ridge,
 	TN 37831-6354; \\
	Department of Physics and Astronomy, University of Tennessee,
	Knoxville, TN 37996-1200; \\ 
	Joint Institute for Heavy Ion Research, Oak Ridge National
	Laboratory, Oak Ridge, TN 37831-6374}
%\author{Anthony Mezzacappa}
%\address{Physics Division, Oak Ridge National Laboratory, Oak Ridge,
% 	TN 37831-6354}
\date{Written contribution to the proceedings of 
``Frontiers of Contemporary Physics--II,'' held March 5-10, 2001 at 
Vanderbilt University, Nashville, Tennessee. The proceedings
of this conference are to be made available at
{\tt http://www.fcp01.vanderbilt.edu/}.}
\maketitle

\begin{abstract}
The depth of our theoretical understanding of neutrino flavor mixing should
match the importance of this phenomenon as a herald of long-awaited
empirical challenges to the standard model of particle physics.
After reviewing the familiar, simplified 
quantum mechanical model and its flaws, I sketch the deeper understanding
of both vacuum and matter-enhanced flavor mixing that is found in the
framework of scattering theory. While the simplified model gives the
``correct answer'' for atmospheric, solar, and accelerator/reactor
neutrino phenomena, I argue that a key insight from the deeper picture will
simplify the treatment of neutrino transport in astrophysical 
environments---supernovae, for example---in which neutrinos play a dynamically
important role. 
\end{abstract}

%\pacs{}

\def\nua{\nu_\alpha}
\def\nub{\nu_\beta}
\def\nuanub{\nu_ \alpha \rightarrow \nu_\beta}
\def\vp{{\bf p}}
\def\vl{{\bf l}}
\def\vk{{\bf k}}
\def\vq{{\bf q}}
\def\vx{{\bf x}}
\def\vy{{\bf y}}
\def\vz{{\bf z}}
\def\vs{{\bf s}}
\def\vr{{\bf r}}
\def\vu{{\bf u}}
\def\bp#1{{\underline p_{#1} }}
\def\bk#1{{\underline k_{#1} }}
\def\enu{E_{\vq}}
\def\gmu{\gamma^\mu}
\def\dmu{\partial_\mu}
\def\lhat{\hat{\bf L}}

%%%%%%%%%%%%%%%%%%%%%%%%%%%%%%%%%%%%%%%%%%%%%%%%%%%%%%%%%%%%%%%%%
\section{Introduction}
\label{sec:intro}
%%%%%%%%%%%%%%%%%%%%%%%%%%%%%%%%%%%%%%%%%%%%%%%%%%%%%%%%%%%%%%%%%

Because the standard model of particle physics admits no neutrino mass, 
any evidence of flavor mixing of massive neutrinos 
challenges this reigning paradigm. 
While the extensions of the standard model that are required
to give mass to the neutrino can be relatively trivial,
patterns of neutrino mass and flavor mixing may provide interesting
clues to the high-energy unification of the separate interactions that
the standard model describes with such precision at currently
accessible energies.

There are various phenomena that beg to be interpreted as neutrino
flavor mixing. 
Some hints and constraints come from observations
of solar neutrinos, reactor and accelerator neutrino experiments, 
and consideration of the role of neutrinos in 
supernovae and Big Bang nucleosynthesis \cite{phenom}.  
Of particular
note is the observation, in the Super-Kamiokande detector,
of large numbers of neutrinos produced in the impacts of cosmic rays 
upon Earth's atmosphere \cite{superk}. These observations establish, with 
high statistics, an asymmetry between the upward and downward $\nu_\mu$ fluxes.
This asymmetry is naturally explained as resulting from the sinusoidal
flux variation with path length expected to result from neutrino flavor mixing.
It may well be remembered in the future as one of the first important
pieces of empirical
evidence of physics beyond the standard model. Long-baseline terrestrial
neutrino experiments will test the flavor mixing interpretation of the
Super-Kamiokande data; initial results from the first of these involving
a $\nu_\mu$ beam (the K2K experiment) show a $\nu_\mu$ deficit 
\cite{k2k}, but more data are needed for a definitive confirmation.

(About three months after this conference, the SNO collaboration announced
an empirical milestone regarding solar neutrinos. 
Comparison of the detected rate of neutrino absorption on deuterium 
(a charged current process) with the rate of elastic scattering of electrons
(which has contributions from both charged and neutral currents) 
provides evidence that a portion of the solar neutrino flux is comprised
of $\nu_\mu$ or $\nu_\tau$ \cite{sno}.)

Our theoretical understanding of neutrino flavor mixing should be 
sufficiently deep and secure to be worthy of its important role
as a phenomenological beacon, signaling towards us from the energetically
distant shores of grand unification. 
To this end, I first review the familiar, 
simplified 
quantum mechanical model. This model embodies the simple heart of the physics,
but also contains flaws that appear under close examination. 
Next I sketch the deeper understanding
of both vacuum and matter-enhanced flavor mixing that is found in
considering neutrino production, propagation, and detection as a
single unified scattering process. In conclusion, I argue that while
the quantitative differences predicted by the deeper framework are  
not discernable in current experiments, a fundamental insight 
from this framework
will be helpful in conceptualizing neutrino transport in astrophysical
phenomena---such as supernovae---in which the free-streaming approximation
breaks down.

%%%%%%%%%%%%%%%%%%%%%%%%%%%%%%%%%%%%%%%%%%%%%%%%%%%%
\section{Simplified Quantum Mechanical Model}
\label{sec:simple}
%%%%%%%%%%%%%%%%%%%%%%%%%%%%%%%%%%%%%%%%%%%%%%%%%%%%

In describing massless neutrinos, the standard model employs only one of 
two spin-1/2 representations of the Lorentz group. These 
``left-handed'' neutrino fields 
annihilate and create neutrino states of negative 
helicity and antineutrino
states of positive helicity. For example, a low-energy effective interaction
responsible for the emission or absorption of neutrinos via interactions
with nucleons 
can be written as a product of leptonic and hadronic currents:
\begin{equation}
{\cal L}_I=-g\; \overline\alpha\,\gamma^\mu(1-\gamma_5)\nu_\alpha\;
\overline p\,\gamma_\mu(1-g_A\gamma_5)n + {\rm H.c.},
\label{interaction}
\end{equation}
where $g$ and $g_A$ are coupling constants, and $\alpha$, $\nu_\alpha$, 
$p$, and $n$ are Dirac fields that respectively 
annihiliate particles and create antiparticles of the following types: 
charged leptons
of flavor $\alpha$; neutrinos associated with those charged
leptons; protons; and neutrons. 
The $(1-\gamma_5)$ factor projects the one needed spin-1/2 representation
of the Lorentz group from the Dirac spinor $\nu_\alpha$. In a description of
massless neutrinos, this so-called left-handed field $\nu_{\alpha L}={1\over2}
(1-\gamma_5)\nu_\alpha$ is all that appears in the free-field part of
the Lagrangian as well.

While empirically undetectable when the standard model was put together,
there is reason to believe that neutrinos have mass. In contrast to
electrodynamics---whose observed gauge invariance requires a vanishing
photon mass---there is no symmetry principle\footnote{Except, perhaps,
lepton universality; see Eq. (\ref{interactionMass}), and the
surrounding discussion and footnote.} to demand a vanishing 
neutrino mass. If one subscribes to the dictum that whatever is not
forbidden is not just allowed, but {\em mandatory}, then neutrino mass
is expected at some level.

A trivial extension to the standard model that allows neutrino mass is
the introduction of right-handed neutrino fields, allowing Dirac mass 
terms.\footnote{Alternatively, an extension to the Higgs sector would allow 
Majorana mass terms.}
The free-field neutrino
Lagrangian then becomes
\begin{equation}
{\cal L}_0=\sum_i \overline\nu_i(i\gamma^\mu\partial_\mu-m_i)\nu_i,
\label{freeLagrangian}
\end{equation}
where the subscript $i$ labels the neutrino masses.
As experience with the weak interactions of the quark sector shows,
the fields $\nu_i$ appearing in the diagonalized free-field Lagrangian
need not be the same as the fields appearing in interactions (e.g.,
Eq. (\ref{interaction})) in which neutrinos are produced in association
with charged leptons. In particular, these fields can be related by
\begin{equation}
\nu_\alpha = \sum_i U_{\alpha i}\,\nu_i,\label{fields}
\end{equation}
where $U_{\alpha i}$ are the elements of a unitary matrix.
In the quark sector, the analogous matrix is known as the 
CKM (Cabbibo-Kobayashi-Maskawa) matrix \cite{ckm}. In the neutrino
case, it is becoming fashionable to call $U$ the MNS
(Maki-Nakagawa-Sakata) matrix \cite{mns}.

To discern the consequences of a nondiagonal MNS matrix $U$,
consider the following quantum mechanical model of neutrinos propagating
in vacuum. (This simplifed model is {\em not} a full quantum field theory of
the neutrino. In particular, the neutrino states defined below are not
obtained by acting on the vacuum with a creation operator obtained from
a quantized neutrino field.) 
Postulate the existence of a Hilbert space, inhabited by the neutrino state, 
whose dimension is equal to the
number of neutrino flavors. The Hamiltonian is simply the free
particle energy. 
Two relevant bases that span this
Hilbert space are (1) a flavor basis $|\nu_\alpha\rangle$, in which
the neutrino is produced and detected; and (2) a mass basis $|\nu_i\rangle$,
which are eigenstates of the Hamiltonian (with eigenvalues $\sqrt{
p^2+m_i^2}$, taking all neutrino states to have the same three-momentum
of magnitude $p$). 
Motivated by Eq. (\ref{fields}), take these two bases to be related by
\begin{equation}
\langle\nu_\alpha |\nu_i\rangle=U_{\alpha i}.\label{overlap}
\end{equation}
The neutrino state $|\Psi(t)\rangle$ at time $t$ is
\begin{equation}
|\Psi(t)\rangle=\sum_i e^{-i\sqrt{p^2+m_i^2}\,t}|\nu_i\rangle\langle \nu_i|
\Psi(0)\rangle.
\end{equation}
Suppose the neutrino was produced at $t=0$
in conjunction with a charged lepton
of flavor $\alpha$; then $|\Psi(0)\rangle=|\nu_\alpha\rangle$. We ask
for the probability at time $t$ for the neutrino state to be $|\nu_\beta
\rangle$. In the relativistic limit, and using Eq. (\ref{overlap}), we find 
\begin{equation}
P_{\nu_\alpha\rightarrow\nu_\beta}(t)=\left|\sum_i U_{\beta i} U_{\alpha i}^*
\exp\left(-i{m_i^2\over 2 p}t\right)\right|^2.
\label{probability}
\end{equation} 
%Hence production by a neutrino beam 
%of charged leptons of a different flavor from those associated
%with the creation of the neutrino beam can result from nonzero neutrino
%masses! 
With nonzero neutrino masses, a neutrino beam produced in conjuction
with charged leptons of flavor $\alpha$ can,
when absorbed in a detector, produce charged leptons of
flavor $\beta$.
This method of detecting neutrino mass is typically much more
sensitive than direct kinematic searches, because large times between
production and detection can overcome small values of $m_i^2/2p$ to 
produce significant phases in Eq. (\ref{probability}). 

Neutrino mixing is altered when the neutrinos pass through matter
\cite{wolf78}.
Coherent forward scattering with amplitude $f$ off of scatterers with number
density $N$ induces an index of refraction $n$, given by the classic formula
\begin{equation}
n-1={2\pi N\over p^2}f.\label{index1}
\end{equation}
(The matter background is assumed to be of sufficiently low density 
that neutrino
absorption and non-forward scattering can be neglected.)
An index of refraction reduces the neutrino velocity below the speed of light;
hence it can also be thought of as an effective squared mass $m_{\rm eff}^2$:
\begin{equation}
n-1={m_{\rm eff}^2\over 2p^2}.\label{index2}
\end{equation}
The effective mass determined by Eqs. (\ref{index1}) and (\ref{index2}) 
is added to the Hamiltonian of the quantum mechanical model of neutrino
evolution. Because the interactions used to compute the forward scattering
amplitude $f$ are flavor-dependent, this piece of the Hamiltonian is
not diagonal in the mass basis. Diagonalization of this new Hamiltonian  
yields new effective (``in-medium'') mass eigenstates and a new mixing matrix. 

This alteration of flavor mixing in the presence of a matter
background can give rise to interesting effects.
Suppression
or enhancement of flavor mixing can occur, depending on the density
of the background, the neutrino momentum, the original (``vacuum'') 
mixing parameters, etc. In particular, the right combination of background
density and neutrino energy induces a resonance, in
which even neutrino flavors that hardly mix in vacuum (small off-diagonal
terms in $U_{\alpha i}$) become maximally mixed \cite{barg80}.
Eventually, it was recognized \cite{mikh86} that the $\nu_e$ produced
in the core of the sun by nuclear burning could encounter such a resonance
while traversing the solar density gradient. The enhanced flavor mixing in 
the resonant region would cause a reduced $\nu_e$ flux emerging from the sun,
complemented by a flux of neutrinos of a different flavor. This is the
Mikheyev-Smirnov-Wolfenstein (MSW) effect.

%%%%%%%%%%%%%%%%%%%%%%%%%%%%%%%%%%%%%%%%%%%%%%%%%%%%%%%%%%%%
\section{Cracks in the foundation}
%%%%%%%%%%%%%%%%%%%%%%%%%%%%%%%%%%%%%%%%%%%%%%%%%%%%%%%%%%%%

Clearly, this quantum mechanical model of flavor mixing has given
a great deal of insight into interesting physical effects. It has
a long and distinguished history, tracing back to the seminal papers on
$K^0-\overline K^0$ physics \cite{kkbar}, from which the notion
of neutrino mixing (or ``oscillations'') 
was suggested by analogy \cite{pont}. The model also has some
flaws, however.

Consider how the mixing probability of Eq.
(\ref{probability}) would be used in describing a neutrino mixing 
experiment. In a
canonical, idealized experiment, 
neutrinos are produced at a source in association
with a charged lepton of flavor $\alpha$, and measured (with perfect
efficiency) with a detector at a distance $L$ from the source
via production of a charged lepton of flavor $\beta$. The event
rate seen in this experimental setup is
\begin{equation}
d\Gamma_{\alpha\beta} = \int dE_{\vq}  \left(d\Gamma_{\alpha,\nua} \over
	L^2\, d\Omega_{\vq} \, dE_{\vq} \right) \left(P_{\nuanub}\right)
	\left(d\sigma_{\nub,\beta}\right), \label{rate}
\end{equation}
where the direction of the neutrino momentum $\vq$ points
from the source to the detector.
The first factor in the integrand represents the
flux of neutrinos of energy $E_{\vq}$ from a process involving a charged lepton
of flavor $\alpha$, and the third factor is the cross section
for neutrino detection via a process involving a charged lepton
of flavor $\beta$. These factors are computed by the standard techniques
of quantum field theory (QFT). In contrast, the middle
factor is computed with the simplified quantum mechanical model
described in the previous section.

Pondering the use of the simplified model's mixing probability in an
experimental event rate like Eq. (\ref{rate}) raises some
conceptual difficulties:
\begin{enumerate}
\item The Hamiltonian evolution describes
the time development of the neutrino state; but usually experiments
are stationary in time, measuring the mixing probability in space.
(Recall that $L$ is the source-detector distance in Eq. (\ref{rate});
the mixing probability as a function of $L$, not $t$, is required.)
\item The dynamics of the spin degree of freedom are not addressed. 
The neutrino states employed in the simplified model of Sec. \ref{sec:simple}
do not carry a spin quantum number. If a spin quantum number were included,
a formula like Eq. (\ref{rate}) would not properly describe interference
between spin states. This is because the parity-violating weak interactions
yield different amplitudes for production and detection of the different
neutrino spin states. To maintain the factorization of 
Eq. (\ref{rate}), at best it could only be written as an incoherent sum
over the contributions of the different spin states. 
\item From a field-theoretic perspective, the whole notion of 
``flavor eigenstates'' is problematic.
The usual procedures for computing neutrino production rates
and cross sections (the first and third factors in Eq. (\ref{rate}))
involve external particles described as plane waves, resulting in
conservation
of both energy and momentum. These procedures force the neutrino
to have a definite mass, a fact of importance to the interpretation
of kinematic searches for neutrino mass \cite{shro80}. 
Instead of using interactions with a charged lepton of flavor
$\alpha$ in the form of Eq. 
(\ref{interaction}), I propose---and this will be an important point
in the discussion of neutrino transport in Sec. \ref{sec:lesson}---that 
such terms be thought of as three separate interactions: 
\begin{equation}
{\cal L}_I=-\sum_i g_{\alpha i}\; 
\overline\alpha\,\gamma^\mu(1-\gamma_5)\nu_i\;
\overline p\,\gamma_\mu(1-g_A\gamma_5)n + {\rm H.c.},
\label{interactionMass}
\end{equation}
where 
\begin{equation}
g_{\alpha i}\equiv g\, U_{\alpha i}\label{mixedCoupling}
\end{equation}
is the coupling constant
governing the production rate and cross section 
of a neutrino of mass $m_i$ in association
with a charged lepton of flavor $\alpha$.\footnote{This formulation
suggests that
the same mechanism that gives neutrinos mass is also responsible for
the breakdown of lepton universality.} 
However, forcing the neutrinos
``on-shell'' in this way precludes the superposition of mass eigenstates
required for coherent flavor mixing. More fundamentally, 
the fact that flavor states do not have a definite energy (not having a
definite mass) creates an algebraic roadblock to attempts to define
creation and annihilation operators for flavor states in the standard
canonical quantization of field theories\footnote{Creation and annihilation
operators for flavor states have been obtained in a nonstandard
approach to the canonical quantization of mixing fermion fields
\cite{blas95}. The approach yields Fock spaces for massive and flavor
neutrinos that are ``unitarily inequivalent representations of the
canonical anticommutation relations,'' and a flavor vacuum with a 
nontrivial condensate structure. Because the scattering approach 
to mixing phenomena
reviewed in Sec. \ref{sec:scattering} has a more 
direct connection to experiment
and is theoretically sound, I do not find the case for the
physical relevance of the
speculative ideas in Refs. \cite{blas95} to be persuasive.} \cite{giun92}. 
If creation and annihilation operators cannot be defined for flavor states,
the invention of a ``flavor basis'' for use in the simplified 
model of Sec. \ref{sec:simple} is physically suspect. 
%\item If we are dealing with Dirac particles, 
%the spin degree of freedom is not explicitly taken 
%into account.\footnote{For Majorana neutrinos, what is usually called
%the ``antineutrino'' is simply the other spin state of the self-conjugate
%neutrino.}
\end{enumerate}

If these difficulties represent cracks in the foundation of neutrino
mixing theory, then the relativistic limit is the cement that fills
the cracks as a temporary fix: 
\begin{enumerate}
\item In the relativistic limit, simply take $t \approx L$.
\item The $V-A$ structure of the neutrino interactions ensures that
in the relativistic limit there is only one relevant spin degree of freedom:
States having the ``wrong'' helicity are projected out by the 
factor $(1-\gamma_5)$ in, for example, interactions like 
Eq. (\ref{interaction}).
\item The production and
detection regions are presumably small enough in size in comparison with the
source-detector distance that, for $m_i \ll p$,
the production and detection regions
contribute negligible phase to Eq. (\ref{probability}). 
The relativistic limit also ensures that the effects of the
small masses can be neglected in phase space factors in the interaction rates.
Taken together, these considerations allow
neutrino mass to be ignored in the production and detection interactions.
Moreover, spatially limited production and detection regions imply
a spread of momentum, allowing sufficient relaxation of energy-momentum
conservation for different mass states to interfere \cite{kays81}.
Finally, the algebraic roadblocks preventing the definition of creation and
annihilation operators for flavor states disappear in the ultra-relativistic
limit.
\end{enumerate}

These explanations may be sufficient to convince one of the basic
physical correctness and quantitative accuracy of the 
simplified quantum mechanical model in the regime of current experimental
relevance (the relativistic limit). Normally, that is enough for 
physicists---especially experimental physicists! Lingering
doubts encroach upon the thoughtful mind, however. 
At first glance, justification of the 
replacement $t\approx L$ might be thought 
of as making a Lorentz transformation from the rest frame of the 
neutrino to the lab frame: The time $t$ in Eq. (\ref{probability})
might be thought of as that
measured by an observer riding with the neutrino between its creation
and detection. 
But in the relativistic limit---so crucial to
the above arguments---there is no rest frame of the neutrino. 
If I then want to
consider the small neutrino mass in making this transformation, which
of the various mass eigenstates' rest frames am I transforming from? 
Presumably they are different: I have chosen equal momenta, but these
momenta correspond to different velocities for different masses.
More importantly, the production flux and detection cross section in
Eq. (\ref{rate}) are not computed in any neutrino rest frame, but in the
lab frame; this makes it obvious that this ``Lorentz transformation'' 
explanation of $t\approx L$ is inexcusible.

The purist, wanting to believe she truly understands the physics, has
a more fundamental question:
How would I proceed if the relativistic limit were
not applicable? For peace of mind, sometimes it is better to tear down 
the foundation and rebuild from the ground up.

%%%%%%%%%%%%%%%%%%%%%%%%%%%%%%%%%%%%%%%%%%%%%%%%%%%%%%%%%%%%%
\section{Scattering Theory Approach to Flavor Mixing}
\label{sec:scattering}
%%%%%%%%%%%%%%%%%%%%%%%%%%%%%%%%%%%%%%%%%%%%%%%%%%%%%%%%%%%%%

A key insight to a more fundamental understanding of flavor
mixing is the recognition that, operationally, one does not
actually prepare or measure neutrino states. Instead, it is
the hadrons and charged leptons that are directly prepared and detected.
The neutrinos are merely {\em intermediate states} in an
overall coherent process of neutrino production, propagation,
and detection \cite{intermediate}. 

In this point of view, an experimental (or observational)
setup for detecting flavor mixing bears some resemblance
to familiar scattering processes treated with Feynman diagrams. 
At low
energies (much smaller than the $W^{\pm}$ and $Z^0$ boson masses)
where effective point interactions like Eq. 
(\ref{interaction}) are applicable, one can think of the
entire neutrino production/propagation/detection process
as a tree level Feynman diagram, with a neutrino propagator
connecting the external hadrons and charged leptons 
involved in the production and detection processes.
One need not worry about the existence of ``flavor eigenstates'',
those ill-defined superpositions of mass eigenstates. 
Instead, the hadronic/charged lepton
initial and final states can be connected by each of 
the (well-defined) neutrino mass eigenstates. Hence the
flavor mixing phenomenon is seen to result from the summation
of entire diagrams having the same initial and final states,
but neutrino propagators of different (definite) mass.
This summation of diagrams obviates the need to introduce
flavor eigenstates, which, as mentioned earlier, are problematic
from a field-theoretic perspective.

Thinking about neutrino flavor mixing in terms of Feynman
diagrams might seem strange because of the macroscopic length
and time scales involved. When is the
last time you thought of a Feynman diagram stretching from
a beam dump to a detector tens or hundreds of kilometers away?
Or worse yet, from the atmosphere on one side of Earth
to a detector on the other side? Or worse still,
from the sun to a detector on Earth? In such circumstances it
is more natural to imagine the production of a ``real''
(as opposed to ``virtual'') neutrino wave
packet, which subsequently travels a macroscopic distance
before being absorbed in a detector. And yet the 
sinusoidal interference terms in Eq. (\ref{probability})---present
in explanations of the atmospheric neutrino problem, and in the
``just-so'' vacuum oscillation solution to the solar neutrino
problem---require just this kind of coherent superposition
over macroscopic (even astronomical) distances.

Not surprisingly, the macroscopic distances in the problem 
produce complications in comparison with the usual use
of Feynman diagrams. In the familiar procedure, the external particles in the
diagrams are taken to be plane waves. The plane wave
scattering amplitude (the $S$-matrix) contains an overall energy-momentum
conserving $\delta$ function. In squaring the $S$-matrix to get a 
probability, one is therefore faced with a squared $\delta$ function.
This can be dealt with by localizing the system in a spacetime 
``box,'' resulting in volume and time factors that can be interpreted
in such a way as to yield event rates, cross sections, etc. 
This procedure is really ``more a mnemonic than a derivation'' 
\cite{wein},
however. 
It is useful because the interactions of interest usually occur 
(in particle colliders, for instance) in a single, small spacetime
volume. It is more convincingly justified, however, by 
a wave packet description (e.g., Ref. \cite{gold}).
 
In describing neutrino flavor mixing as an overall 
production/propagation/detection process, it is not possible to 
compute event rates
directly from the plane wave $S$-matrix with the usual mnemonic described
above. This is because 
a neutrino oscillation experiment involves neutrino
production and detection regions which are widely separated in
space. In contrast to the case of accelerator particle collisions,
the interactions of interest do not all occur in a
single volume element. In addition, as also noted above,
in this microscopic picture the production and detection of a single
neutrino will be separated in time as well as space.
In order to describe the spacetime localization 
one must fall back on a wave packet description
of the external particles,
in which the amplitude is a superposition of plane wave amplitudes.

Justification of the use of the simplified quantum mechanical model
of Sec. \ref{sec:simple}
to compute an oscillation probability in general cases involves 
the following procedure:
\begin{enumerate}
\item Write down a properly normalized superposition of plane wave $S$-matrix
amplitudes corresponding to wave packets of hadrons and charged leptons 
overlapping in regions of limited extent in space and time in the 
source and detector.
\item Use this amplitude to determine a normalized event rate for the
observation of the appropriate external particles in the source and
detector.
\item Compare this event rate with Eq. (\ref{rate}), the kind of expression
usually used to predict event rates in neutrino mixing experiments. 
If one can identify factors corresponding to the production flux and
detection cross section (the first and third factors in Eq. (\ref{rate}),
then everything else is the oscillation probability.  
\end{enumerate}
While a number of authors \cite{qft} followed the seminal work of Refs. 
\cite{intermediate} in studying aspects of flavor mixing,  
this full three-step procedure was set forth,   
carried to completion, and applied to neutrinos propagating through
matter as well as vacuum in Refs. \cite{card01,card00}.
The procedure shows how the mixing probability obtained from the simplified
model arises in a more rigorous and physically complete way, and also
shows how the interference terms are gradually lost. As a bonus, one
sees how the phenomenon would work in nonrelativistic cases, and
spin is automatically taken into account through use of the 
neutrino propagator.
In the relativistic limit, factorization into the form
of Eq. (\ref{rate}) occurs. 

After stripping away the flux and
cross section, what remains is the oscillation probability:
{
\begin{eqnarray}
& &\nonumber\\
P_{\nuanub}\left(\enu,L\right)& =& 
\sum_{k,k'} U_{\alpha k}U_{\beta k}^* U_{\alpha k'}^* U_{\beta k'} 
\nonumber\\ & &\times
\exp \left[- i {(m_k^2 - m_{k'}^2)L\over 2\enu}
-{(m_k^2 - m_{k'}^2)^2 L^2\over 32 \enu^4 \ell^2}
\right]\label{oscprob}.
\end{eqnarray}
}
Here $L$ is the source-detector distance, which arises naturally
from the coordinate space neutrino propagator; unlike the simplified
model, no transformation from time to space is necessary. The 
quantity $\ell$ is a length scale describing the extent in space
and time of the regions of wave packet overlap in the source and
detector. The exponential damping factor does not appear in
the simplified model of Sec. \ref{sec:simple} (cf. Eq. (\ref{probability}));
it arises because the wave packet treatment of external particles
gives rise to intermediate neutrino wave packets which, having different
velocities, begin to separate as the source-detector distance is traversed.
This damping of interference terms probably only comes into play
when $L$ is much larger than the period of flavor oscillations,
a case in which binning over energy and source/detector positions
tends to wash out the interference terms anyway. Still, it is nice
to see that the intrinsic decoherence arises naturally. Finally, the  
expression from the full calculation of Ref. \cite{card01}
corresponding to Eq. (\ref{rate})
automatically contains the time delay between neutrino emission and
detection, an effect that would have to be added by hand when using 
the simplified model. 

A similar calculation has been performed in the case of neutrinos
propagating through a matter background \cite{card00}. 
A goal of this work was to justify the use
of the Schr\"odinger equation of the simplified model, in which
the (varying) effective mass induced by neutrino forward scattering 
is included
in the Hamiltonian. 
%In applications, the flavor mixing probabilities 
%obtained from the simplified model with a matter background are  
%generally averaged; the oscillatory interference terms are ignored,
%being undetectable due to integrations over energy and source position.
%If one is not interested in studying the damping of the interference terms, a 
%stationary calculation can be performed.
%
Unlike the vacuum case---where the explicit vacuum neutrino
propagator was exploited
from the outset---an explicit propagator was not initially assumed in
the calculation of Ref. \cite{card00}.
Instead, it was first determined what the general form 
of the coordinate space neutrino propagator
% (actually, a mixed form
%of the propagator that depends on the spatial coordinates and the energy)
must be
in order for the factorization of Eq. (\ref{rate}) to occur.
%The case of a uniform background was studied first. It was found that
%the mass degeneracy of the spin states of the neutrino was lifted.
%In the relativistic limit, only states of the correct spin are
%projected out. The propagator then takes on the form necessary
%for factorization of Eq. (\ref{rate}), with the factor constituting
%the flavor mixing amplitude taking on the form expected from the simplified
%model (but with spatial instead of time dependence). 
%Guided by the
Guided by this
expected form of the propagator, the case of a spatially varying background
was 
%also 
studied in the relativistic limit, and it was shown that
the factor from the neutrino propagator 
giving rise to the mixing amplitude does in fact obey
(a spatial version of) the Schr\"odinger equation employed in the
simplified model of Sec. \ref{sec:simple}.

%%%%%%%%%%%%%%%%%%%%%%%%%%%%%%%%%%%%%%%%%%%%%%%%%%%%%%%%
\section{A Lesson To Take Away}
\label{sec:lesson}
%%%%%%%%%%%%%%%%%%%%%%%%%%%%%%%%%%%%%%%%%%%%%%%%%%%%%%%%

While the simplified quantum mechanical model of neutrino
flavor mixing phenomena reviewed in Sec. \ref{sec:simple}
captures the heart of the effect
under special (though experimentally relevant) conditions,
the scattering-process approach to neutrino flavor mixing
phenomena is physically realistic. The fact that it shows
how to proceed even in nonrelativistic situations---or with
more general interactions than the usual $V-A$ case---shows
that a deeper understanding has been attained. One now
appreciates that a separate, simplified model of neutrino
flavor mixing is a special situation; consideration of the
more general picture provides the realization that flavor
mixing is a coherent process in which the mechanisms of production and
detection of the ``particle mixture'' are entangled. That
Nature should make the oscillation length of two of her
neutrino flavors coincide with the diameter of Earth is already
startling; to realize now that Super-Kamiokande is observing
a coherent superposition of Feynman diagrams the size of Earth's
diameter
is truly amazing!

Beyond the satisfaction of understanding, however,
lies a lesson of practical use in the way neutrino
transport might be treated in astrophysical phenomena such as
supernovae. 

The simplified quantum mechanical model described in Sec.
\ref{sec:simple} assumes free streaming (or free streaming + coherent
forward scattering) neutrinos, but in astrophysical situations
interaction rates may exceed or compete with the flavor oscillation period.
Neutrino scattering, production, and absorption events are 
incoherent processes that interrupt coherent flavor oscillations
and drive the system to chemical (flavor) equilibrium.
A density matrix, with its ability to describe partial coherence, 
is a natural construct to employ in describing
the combined effects of flavor oscillations and interactions. 
%Such an approach has been developed over the years 
%that employs a density matrix assembled from the flavor/mass
%eigenstates of the simplified model \cite{raff96}. 
Such an approach has been developed over the years 
(\cite{density,offdiagonal}; for a review see Ref. \cite{raff96}),
in which the quantum state vectors used to construct a density matrix 
are the flavor/mass
eigenstates of the simplified quantum model of Sec. \ref{sec:simple}. 
This approach
works in the relativistic limit, and has been used to study
effects of flavor mixing during Big Bang nucleosynthesis, in
the cores of supernovae, and in the tenuous wind environment
outside a nascent neutron star. Each of these environments
is either homogeneous or quasi-stationary, and this suits them
to a treatment (the density matrix approach of Refs. 
\cite{density,offdiagonal,raff96}) 
deriving from the simplified model of Sec. {\ref{sec:simple}, which 
allows evolution in time {\em or} space to be followed. However, it
is not clear how or if this approach could be adapted
for use in, say, a dynamic supernova simulation, in which variation
in time {\em and} space must be followed.

In the brave new world of empirically confirmed neutrino mass, 
the question of how to cope with neutrino transport with flavor mixing 
acquires a new urgency.
Is there a way to
handle neutrino 
transport---{\em with variation in space
and time}---with classical methods (i.e. the Boltzmann equation),
while still capturing ``flavor mixing'' physics?
In responding to this question, we should remember an important lesson
from the more in-depth study of flavor mixing physics outlined
above: {\em Fundamentally, there is no such thing as flavor eigenstates.}
I will argue below that under certain conditions that 
make a classical treatment feasible,
we should consider banishing 
talk of ``electron neutrinos,'' ``muon neutrinos,''
and ``tau neutrinos;'' instead, we should perhaps speak only of
``$m_1$ neutrinos,''
``$m_2$ neutrinos,'' ``$m_3$ neutrinos,'' etc.\footnote{These neutrinos
of definite mass should be given more interesting names.
Perhaps they could be called $\nu_m$, $\nu_n$, and $\nu_s$, where the letters
come from the authors \cite{mns} for whom the neutrino mixing matrix
is named; or the Brahma, Vishnu, and Shiva neutrinos; or something else
sufficiently different from $\nu_e$, $\nu_\mu$, and $\nu_\tau$.}
While the seminal papers on $K^0-\overline K^0$ physics \cite{kkbar}
introduced the use of a simplified quantum mechanical model for 
``particle mixtures,'' Gell-Mann and Pais were also sensitive to the problem of
what the ``true particles'' are. They noted that 
an attempt to introduce ``quanta''
for objects not obeying
fundamental conservation laws ``can only be a mathematical device that
distracts our attention from the truly physical particles.'' They pointed
out that the word ``particle'' should be reserved for the objects having
a definite lifetime and mass. In the case of neutrinos, ``flavor
states'' are not Lorentz invariant, not having a definite mass.  
Following the advice of Gell-Mann and Pais, ``flavor states'' should 
not be thought of as true particles; in fact, attempts to define
``quanta'' (creation and annihilation operators) for them run into trouble
\cite{giun92}. Classical transport can only deal with ``true particles,''
not inherently quantum objects like ``particle mixtures.''
Therefore, I now investigate the possibility that 
classical transport of neutrinos with mass
should, under certain conditions, be done in terms of the ``mass eigenstates.''

Let us examine why thinking only in terms of the mass eigenstates---the 
valid ``quanta'' or ``truly physical particles''---might 
allow for a classical treatment
of transport while still capturing some ``flavor mixing'' physics. 
Consider the behavior of the
oscillation probability, Eq. (\ref{probability}) or Eq. (\ref{oscprob}), 
as the neutrino
flight time $t$ or distance $L$ becomes large in comparison with the
``oscillation
lengths'' associated with the various neutrino squared mass differences, 
$l^{\rm osc}_{i j}\equiv (2\pi E_\vq)/(m_i^2-m_j^2)$. 
In this limit the interference
terms are washed away (whether due to intrinsic decoherence---the exponential
damping in Eq. (\ref{oscprob})---or integration of many oscillations over
a finite energy range), so that
\begin{equation}
P_{\nuanub}\rightarrow\sum_i|U_{\alpha i}|^2|U_{\beta i}|^2.
\label{averageProbability}
\end{equation}
Inserting this mixing probability into the experimental event rate
of Eq. (\ref{rate}), we see that this event rate can now be expressed 
\begin{equation}
d\Gamma_{\alpha\beta} = \sum_i \int dE_{\vq} \left(d\Gamma_{\alpha,\nu_i} \over
	L^2\, d\Omega_{\vq} \, dE_{\vq} \right)
	\left(d\sigma_{\nu_i,\beta}\right). \label{classicalRate}
\end{equation}
%where the two factors of the integrand now represent a neutrino 
%production flux
%and cross section computed (using standard plane wave methods) 
%with interactions like 
%Eq. (\ref{interactionMass}), in which the mixing parameters have been 
%absorbed into the coupling constants. 
%This makes perfect sense in light
%of the discussion surrounding Eq. (\ref{interactionMass}). 
In arriving at Eq. (\ref{classicalRate}), the factors $|U_{\alpha i}|^2$ 
and $|U_{\beta i}|^2$ have respectively been absorbed into the production
flux and detection cross section, and the sum over intermediate neutrino
mass
states taken outside their product. The experimental rate to detect
different flavors of charged leptons at the source and detector is no 
longer a single ``flavored neutrino'' flux, multiplied by an oscillation 
probability, multiplied by a 
``flavored neutrino'' cross section, as in Eq. (\ref{rate}); 
it is now a sum of several contributions, each of which is a
``massive neutrino'' flux times a ``massive neutrino'' cross section. 
The ``flavored neutrino'' fluxes and cross sections in Eq. (\ref{rate})
are computed 
(using standard plane wave methods) with
interactions like Eq. (\ref{interaction});
in contrast, the massive neutrino fluxes and cross sections 
appearing in Eq. (\ref{classicalRate}) are computed
with interactions like 
Eq. (\ref{interactionMass}), in which the mixing parameters $U_{\alpha i}$
have been 
absorbed into the coupling constants (see Eq. (\ref{mixedCoupling})).

As expected, with no interference
terms, the situation can be expressed classically---as an incoherent sum.
Notice, however, that the ``flavor mixing'' is still there: Different
flavors of charged leptons ($\alpha$ and $\beta$) are associated
with production and detection. In this incoherent limit, the discussion
surrounding Eq. (\ref{interactionMass}) comes into play. The neutrinos
are indeed forced ``on shell'' at production and detection, but
each of these massive neutrino types has a coupling (given by Eq. 
(\ref{mixedCoupling})) to each type of charged lepton, so that
(an incoherent version of) flavor mixing still operates.
Kinematically---that is, for the purpose
of simplifying phase space factors in the rates---the relativistic limit
can still be taken. The production rates and cross sections in Eq. 
(\ref{classicalRate}) will then be the same as the massless case,
except for a difference in coupling constants. {\em But even if the mass can
be neglected for phase space purposes, remembering that the ``true
particles'' are the mass states makes all the difference in arriving
at an experimental rate that retains flavor mixing effects in the
classical limit.}     

The validity of using the classical 
``flavor mixing'' rate, Eq. (\ref{classicalRate}), may be challenged:
It appears to conflict with the landmark experiment at Brookhaven
in 1962 \cite{danb62}
interpreted 
as establishing the existence of the $\nu_\mu$ as a separate neutrino species.
Neutrinos from pion decay (with an associated antimuon) were observed 
to produce muons in a detector, and not electrons. 
In principle there are two possible reasons for this. First, it may be
that Eqs. (\ref{averageProbability}) and (\ref{classicalRate}) are applicable,
but that the MNS matrix $U_{\alpha i}$ 
is very close to diagonal. In this case, one neutrino mass state
would have a strong coupling to the muon (see Eq. 
(\ref{mixedCoupling})) and be produced in abundance; but it would have
a small coupling to the electron, such that the rate of appearance of
electrons in the detector would be below experimental sensitivity. 
While one of the other neutrino 
mass states would have a strong coupling to the electron, this state
would only be weakly coupled to the muon, so that too few of these
neutrinos would be produced
in the source to cause a noticeable electron appearance rate in the
detector.
%that the experiment was
%not sufficiently sensitive to detect the contributions arising from
%off-diagonal terms. 

A second possible interpretation of the landmark
Brookhaven experiment is that 
that Eqs. (\ref{averageProbability}) 
and (\ref{classicalRate}) are not applicable; this would be the case
if the source-detector distance $L$ is much smaller than the ``oscillation
lengths'' $l^{\rm osc}_{ij}$ 
associated with the various neutrino squared mass differences. 
In this case, the mixing probability
reduces to
\begin{equation}
P_{\nuanub}\rightarrow \delta_{\alpha\beta},
\label{diagonalProbability}
\end{equation}
so that the experimental flavor mixing rate becomes
\begin{eqnarray}
d\Gamma_{\alpha\beta}& =& \int dE_{\vq} \left(d\Gamma_{\alpha,\nua} \over
	L^2\, d\Omega_{\vq} \, dE_{\vq} \right)
	\left(d\sigma_{\nua,\alpha}\right),\ \ \ \ \ (\alpha=\beta)
\nonumber \\
	&=& 0,\ \ \ \ \ \ \ \ \ \ \ \ \ \ \ \ \ \ \ \ \ \ \ \ \ 
	\ \ \ \ \ \ \ \ \ \ \ \ \ \ \ \ \ \  (\alpha\ne\beta),
\label{diagonalRate}
\end{eqnarray}
where the ``flavored neutrino'' production flux and detection cross section
are computed with interactions like Eq. (\ref{interaction}).
%a possibility also consistent with experiment.  
Current phenomenology \cite{phenom}
of both atmospheric and solar neutrinos suggests that this 
second possibility is the correct interpretation.
Both favor strong mixing (large off-diagonal terms in $U_{\alpha i}$), 
which would have
caused observable flavor mixing if Eqs. (\ref{averageProbability}) 
and (\ref{classicalRate}) were applicable.
Atmospheric and solar neutrino data also indicate that 
neutrino squared mass differences
are too small
to have been probed by the source-detector 
distance in the Brookhaven experiment.

The application of
these considerations to neutrino transport comes from
a correspondence between a flavor mixing experiment
and neutrino processes in an
astrophysical environment:
Microscopic neutrino emission,
absorption, and scattering events in an astrophysical
environment are analogous to the
processes occuring in the neutrino production and detection
regions of a flavor mixing experiment, while the neutrino
mean free path in an astrophysical setting corresponds to
the experiment's source-detector distance.

Based on this correspondence, one concludes that a classical (Boltzmann 
equation) treatment is possible if the neutrino mean free path 
is either much longer or much shorter than the oscillation length.
The long mean free path case involves complete incoherence, 
entailing the physics 
discussed in connection with Eqs. (\ref{averageProbability}) 
and (\ref{classicalRate}). In particular, 
with long mean free paths---{\em and in the absence of coherent 
forward scattering effects}---one would think in terms of the neutrino 
vacuum mass
states, and rewrite all interactions in the manner of Eq. 
(\ref{interactionMass}), with the mixing matrix elements $U_{\alpha i}$
absorbed into
the coupling constants, as in Eq. (\ref{mixedCoupling}).
These revised interactions would then enter into the emissivities
and opacities appearing the Boltzmann equation.
On the other hand, the short mean free path case involves full
coherence, but without time for phase differences between the contributing
intermediate states to be established, as discussed in connection
with Eqs. (\ref{diagonalProbability}) and (\ref{diagonalRate}).
For short mean free paths, one can 
pretend that the neutrinos are massless ``flavor states,'' 
and employ the standard electroweak interactions to derive emissivities
and opacities.

When effective contributions to neutrino effective mass from
forward scattering off a background medium are considered, there
are two complications to the proposed classical treatment in the
fully incoherent (long mean free path) limit. 
First, the relationship between flavor fields and the ``true particle''
quanta is not as simple as in the vacuum case, which involves
an overall matrix relation (i.e., Eq. (\ref{fields})
between flavor fields and ``mass eigenstate
in matter'' fields. Second, a medium with varying
density---in which resonant enhancement of flavor mixing can occur,
as discussed in Sec. \ref{sec:simple}---introduces a new length (and/or
time) scale into the problem.

When background matter effects are included in the Hamiltonian
of the simplified model of Sec. \ref{sec:simple}, 
the newly diagonalized Hamiltonian
yields ``mass eigenstates in matter'' that differ from the vacuum
mass eigenstates. 
These new eigenstates should now be considered the ``true particles,''
playing the same role the vacuum mass eigenstates played in the
previous discussion of the incoherent (long mean free path) 
case.\footnote{This borrows from the point of
view of condensed matter physics, in which describing
systems in terms of ``quasi-particles'' often provides 
tremendous simplification. In describing the observation
of a new class of excitations called ``orbitons,'' it was
noted that ``The prefix quasi is unnecessary. These particles 
are as real as the more familiar photons,
electrons and protons'' \cite{alle01}.}
As discussed previously, if the oscillation length
is short compared with the mean free path, a classical 
(Boltzmann) treatment of transport may be justified, where now 
distribution functions are defined for the ``mass eigenstates in matter.''
However, a complication arises due to the fact that the effective 
mass contribution of 
the matter background depends on the neutrino momentum. 
This causes the new ``in-matter'' mixing matrix (the unitary matrix
that diagonalizes the Hamiltonian of the simplified model) 
to depend on the neutrino
momentum as well, so that there is no overall matrix relation
like Eq. (\ref{fields}) between flavor fields and ``mass-eigenstate
in-matter'' fields. In a deeper picture, 
 quantization of the neutrino field
in a background can still be accomplished
\cite{mann88}, though the mixing matrix must now reside inside the
integral over the momentum variable of the quantized neutrino field.

By rewriting interactions like Eq. (\ref{interaction})
in terms of these
quantized neutrino-in-matter fields, 
it should still be possible to redefine the emissivities and
opacities to describe the emission and absorption processes
in terms of the new ``true
particles.'' In particular---and this is a crucial point, 
as discussed below---{\em the 
coupling constants describing
the interaction strength between the ``mass-in-medium'' neutrino
states and the charged leptons will be determined by the ``in-medium''
mixing matrix, in a manner conceptually similar to Eq. 
(\ref{mixedCoupling}).}

Another complication concerns the fact that in the supernova
environment the matter background varies in space and time.
To deal with this, one can consider a coarse graining into macroscopically
small but microscopically large elements, each with a
constant matter background.\footnote{Such a
coarse graining is analogous to the way one
develops a concept of local thermodynamic equilibrium.
In the context of a computer simulation it would be natural
to identify each coarse-grained 
element with a zone of the discretized system.}
Each coarse-grained element, having a different background density, 
will have its own values of the in-medium mass eigenvalues. However, 
if the variation of the background from one element to another
(and one time step to the next) is small compared with the
local oscillation length (and time), one can 
(along the lines of Ref. \cite{bethe86})
consider the ordered mass eigenvalues in adjacent elements
to correspond to the same ``true particle'' types.\footnote{The
validity of this approach in regions 
of rapidly varying background in supernovae---such as the 
``density cliff'' at the surface of the nascent 
proto-neutron star, and the shock created during the ``bounce''
of the collapsing core---will have to be examined in more detail.}

Hence, in a Boltzmann treatment in the incoherent (long mean free path)
limit,
one would follow the distribution functions of the highest mass
in-medium neutrino, that of next-highest mass, etc. Even though
these masses are slowly changing, one can still consider that one is
following the ``true particles.'' 

Note that this procedure automatically
reproduces the
the MSW effect. The key to seeing this is to note that,
as mentioned above, the coupling strengths of the
mass-in-medium neutrinos to the charged leptons are determined
by the in-medium mixing matrix. This mixing matrix---which diagonalizes
the effective neutrino mass matrix including background matter 
effects---will vary with changing density, leading to changes in the relative
coupling strengths of the in-medium mass eigenstates.  
Therefore, the MSW ``flavor transformation''
is manifest in the changes in the relative strengths of coupling 
of a given in-medium definite mass neutrino type to
the various charged leptons at different positions and times. 
These local coupling constants will be incorporated into the emissivities
and opacities of the in-medium neutrino species, so that evolution
of the Boltzmann equation automatically tracks to propensity of the
distribution functions of these ``true particles'' to interact with
the various flavors of charged leptons. 

The ``adiabatic'' treatment---in which the local oscillation length
(and time) is assumed small in comparison with the scale of variations 
(in space and time) in the background---can break down when two different
mass eigenvalues approach each other as the background varies. 
%For
%nearly diagonal vacuum mixing matrices, 
In particular,
there can be a large probability
for a neutrino of one mass eigenvalue to ``jump'' or tunnel to the
adjacent mass eigenvalue. This can be treated as a classic level-crossing
problem \cite{haxt86}, 
and the resulting ``jump probability'' might be incorporated
into the Boltzmann equation as an ad-hoc interaction transferring one
kind of ``true particle'' (say, in-medium mass eigenstate 1) into another
``true particle'' type (say, in-medium mass eigenstate 2). 
%However, this may
%not even be necessary if the current phenomenology favoring strong
%(vacuum) mixing is generic, as this suppresses the ``jump probability.''

Focus on the ``true particles''---the in-medium mass eigenstates---has
a subsidiary benefit in simplifying the treatment of the contribution
of background neutrino populations to the neutrino effective mass. 
It has been thought necessary to treat this effect with a density
matrix, because this interaction appears to have off-diagonal 
contributions in the ``flavor basis'' (e.g. Refs. \cite{offdiagonal,raff96}).
However, it was shown in
Ref. \cite{card00} that in the basis of the ``true particles''
this contribution could be expressed in terms of a macroscopic 
neutrino number current.\footnote{That work assumed---erroneously, I would now
argue---that the vacuum mass eigenstates were the relevant
``true particles.'' But the basic result should stand,
that the contribution of the neutrino
background is expressible as a macroscopic current when the right basis
is used.}
(In hindsight, it makes sense that the contribution from the
``true particles'' should be expressible in terms of a classical object!)
Hence, being analogous to the background contribution of, say, 
electrons, treatment of the neutrino background does not by 
itself require a density matrix approach. 

In summary, the lesson to take away is this: The realization from
a deeper study of flavor mixing physics that
there is really no such thing as ``flavor eigenstates'' may have
practical utility in environments like supernovae, where neutrino
transport (ranging from diffusion to free-streaming) must be dealt
with in space {\em and} time. 
We do not usually need interference terms in astrophysical
environments. Needing only classical probabilities, we can hope for
a classical Boltzmann equation treatment. 
When the neutrino mean free path is much smaller than the oscillation
lengths, the neutrinos can be treated as massless ``flavor states,'' 
with emissivities
and opacities derived from the standard electroweak interactions. 
For the case of a mean free path much longer than the oscillation lengths,
I have argued that
a classical Boltzmann treatment is still possible 
{\em and conceptually simple}---while 
still retaining ``flavor mixing'' 
physics---when the distribution functions are taken to
describe the ``true particles,''
the mass eigenstates (vacuum or in-medium as appropriate), and the interactions
are rewritten to absorb the mixing matrix elements into local coupling 
constants.\footnote{The reduction to a Boltzmann-type evolution
was appreciated in Ref. \cite{volk00}, but in my opinion
the continuing emphasis on
the ``flavor basis'' introduces considerable conceptual complication.
Moreover, it is not clear that Ref. \cite{volk00} arrives 
at something that can deal
with transport in space {\em and} time.} 
%This approach will succeed if the oscillation length (and time)
%in matter is much smaller than the neutrino mean free path 
%(and collision time). 

The applicability of these ideas to neutrino transport in supernovae
remains to be explored in detail. For ``sterile'' neutrinos
with masses in the keV range (a warm dark matter candidate \cite{sterile}), 
the long mean free path limit may apply in all regions and at all times
in the supernova environment. For the smaller 
mass differences of ``active'' neutrinos
indicated by atmospheric and solar neutrino data, the short mean free
path limit will obtain in the core, while the long mean free path limit
may be realized further out in the envelope. 
The the viable Boltzmann treatments in these two
limits are conceptually different, and a prescription for piecing them 
together would have to be determined. Ultimately, it may be found that  
a more fundamental approach to transport with mixing is necessary to deal
with such an intermediate region, 
at least to determine a valid matching prescription.\footnote{More 
fundamental approaches to quantum kinetics
might take the form of an extension of the 
already-developed density matrix approach \cite{density,offdiagonal,raff96}, 
or application of non-equilibrium
field theory via Wigner functions \cite{sireraperez} or Green functions
\cite{yama00}.}
In any case, there remains much exciting
work in fleshing out this proposed perspective on neutrino
transport, and determining its
applicability to active-active
and active-sterile mixing scenarios that might affect supernova
explosions, heavy-element nucleosynthesis, and supernova
neutrino signals
in terrestrial detectors.

%%%%%%%%%%%%%%%%%%%%%%%%%%%%%%%%%%%%%%%
\acknowledgements{I express appreciation to Anthony Mezzacappa, whose careful
reading of the manuscript produced invaluable suggestions for clarifying
the presentation of the ideas in this work. Discussions with
Gail McLaughlin and Irina Mocioiu are
gratefully acknowledged.
I thank  
Daniel J. H. Chung for enjoyable and instructive
collaboration on a portion of the work discussed here.
I am supported by a DoE PECASE award to Anthony Mezzacappa
at Oak Ridge National Laboratory.}
%%%%%%%%%%%%%%%%%%%%%%%%%%%%%%%%%%%%%%%

%%%%%%%%%%%%%%%%%%%%%%%%%%%%%%%%%%%%%%%

\end{document}